\documentclass[twocolumn,secnumarabic, graphics,floatfix,pre]{revtex4-1}

\usepackage{amsfonts}
\usepackage[T1]{fontenc}
\usepackage[ansinew]{inputenc}
\usepackage{graphicx} 
\usepackage{amsmath}
\usepackage{latexsym}
\usepackage{amssymb}
\usepackage{color}
\usepackage[normalem]{ulem}
\usepackage{epstopdf}
\usepackage{float}
\usepackage{textcomp} 

\newcommand{\epn}{{\varepsilon'_n}}
\newcommand{\tnd}{\text{tn}\delta}
\newcommand{\dgam}{\dot \gamma}
\newcommand{\DSs}{$\text{DS}_s$}
\newcommand{\DST}{$\text{DS}_T$}
\newcommand{\epinf}{\varepsilon_\infty}
\newcommand{\eps}{\varepsilon}
\newcommand{\ome}{\omega}

\definecolor{linkcolor}{rgb}{0,0,0.6} 
\usepackage[pdftex,colorlinks=true, pdfstartview=FitV, linkcolor= linkcolor, citecolor= linkcolor, urlcolor= linkcolor, hyperindex=true,hyperfigures=true]{hyperref} 

\begin{document}

\title{Relaxation time of a polymer glass stretched at very large strains}
\author{R. Sahli$^{1,*}$, J. Hem$^1$,   C.  Crauste-Thibierge$^1$, F. Cl\'ement$^2$, D.R.  Long$^2$, S. Ciliberto$^1$,}
\affiliation{$1$ Univ  Lyon, Ens de Lyon, Univ Claude Bernard, CNRS, Laboratoire de Physique, UMR 5672, F-69342 Lyon, France}
\affiliation{$2$ Laboratoire Polym\`eres  et Mat\'eriaux Avanc\'es, CNRS/Solvay, UMR 5268, 87 avenue des Fr\`eres  Perret, 69192 Saint Fons Cedex, France}

\begin{abstract}
	The polymer relaxation dynamic of a sample, stretched up to the stress hardening regime,  is measured, at room temperature,  as a function of the strain $\lambda$  for a wide range of the strain rate $\dot \gamma $, by an original dielectric spectroscopy set up.  { The mechanical stress modifies the shape of the dielectric spectra mainly because it affects the dominant  polymer relaxation time $\tau$, which depends on $\lambda$ and  is a decreasing function of $\dot \gamma$. The  fastest dynamics is not reached at yield but in the softening regime. The dynamics slows down during the hardening, with a progressive increase of $\tau$. A small influence of $\dot\gamma$ and $\lambda$ on the { relative dielectric strength} cannot be excluded. } 
\end{abstract}
\maketitle

\begin{figure}[h]
	\centering  
	\includegraphics[width=0.8\linewidth]{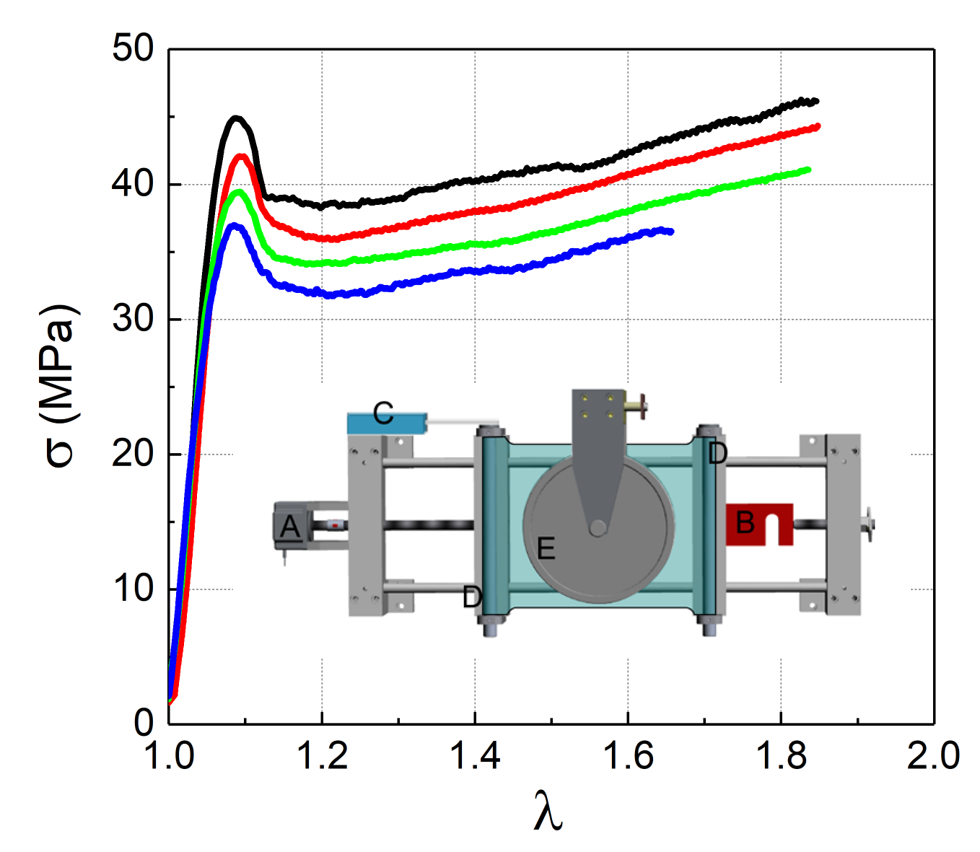}
	\caption{Stress evolution over a wide range of $\lambda$ at several constant strain rates : $2.5\times10^{-3} s^{-1}$ (black), $2.5\times10^{-4} s^{-1}$ (red), $2.5\times10^{-5} s^{-1}$ (green) and $2.5\times10^{-6} s^{-1}$ (blue). Several identical specimens have been measured for each strain rate. 
		{\bf Inset}:  tensile machine for dielectric measurements under stress : motor (A), load cell (B), linear transducer (C), sample fastening cylinders with the stretched polymer film (D) and electrodes (E) connected to the dielectric spectrometer (not sketched).
	}
	\label{Fig:SetUp}
\end{figure}

Mechanical and dynamical properties of polymers are intensively studied, due to their 
fundamental and technological importance~\cite{Ferry}. When strained at a given strain rate, beyond the elastic regime in which the stress is proportional to strain, 
glassy polymers exhibit a maximum in the stress-strain curves (yield point) at a strain of a few percents~\cite{landel} and the deformation becomes irreversible (see Fig.\ref{Fig:SetUp}). At larger strains, depending on the history of the 
sample~\cite{Meijer2003}, the stress drops 
(strain-softening regime) before reaching  a plateau corresponding to plastic flow. Strain-hardening  may then occur at even larger strains, 
depending on the molecular weight and on the cross-linking of the polymer~\cite{meijer2005}. 
The key new insights obtained either by numerical simulations~\cite{hoy2006b,hoy2010,hoy2011} or experiments~\cite{vanmelick2003,govaert2008,senden2010,jatin2014} are that strain hardening 
appears to be controlled by the same mechanisms that control plastic flow 
\cite{chui1999,hasan1993,robbins2009,ge2010}.  However the microscopic mechanisms leading to such a mechanical behavior are not fully understood~\cite{kramer2005,hoy2016}. For example  it is unclear to what extent   the  relaxation  dynamics  in polymer glasses is modified when the sample is stretched  into the plastic region (see for example  refs.\cite{rev_mod_phys_2018,conca2017,Procaccia2016,roth2016}). 
 
The purpose of this letter is to bring new insight into this problem by presenting the results of experiments in which we performed dielectric spectroscopy of polymer samples stretched  till the  strain hardening regime. 
Dielectric spectroscopy, 
allows the investigation of the dynamics of relaxation processes by means of the polarization of molecular dipoles. { It is directly sensitive to polymer mobility and probes directly the segmental motion}. It can be used to quantify the mobile fraction of polymers. {Measuring the dielectric response of polymers \emph{in situ} had been pioneered by Venkataswamy \textsl{et al.} \cite{venkataswamy1982}}.
It is complementary  to other techniques, such as Nuclear Magnetic Resonance  \cite{loo2000} and the diffusion of probe molecules  \cite{ediger2000,bending2014,hebert2015}, used  to study the molecular dynamics of polymers under stress. The dielectric spectroscopy has already been used in combination with mechanical deformation to study the dynamics in the amorphous phase of polymer under stress  \cite{kalfus2012,perez2016dielectric}.
The results of these experiments were limited to the yield point { whereas} the experimental studies of  the microscopic behavior and processes during strain hardening are more scarce. 

In this letter we present the results obtained by our original   experimental apparatus which  can measure with high accuracy the evolution of the Dielectric Spectrum(DS) of a sample stretched, till the strain hardening regime, at different  strain  rates.  Thus we can  precisely compare the \DSs obtained as a function of stress at room temperature  with those (named  \DST) obtained as a function of temperature in an unstressed sample. This comparison allows us to extract useful informations on the relaxation dynamics under stress at various strain rates.  Our experimental  results bring new important informations because they  clearly show an  acceleration of the  dynamics   which reaches a maximum   in the softening regions. Instead the molecular mobility  slows down again during the strain hardening regimes. 
 
Our experimental setup is composed by a home-made dielectric spectrometer coupled with a tensile machine for uniaxial stretching of films~\cite{perez2015simultaneous, perez2016dielectric}. The scheme of the experimental setup is presented in the inset of  Fig.~\ref{Fig:SetUp}. { More details are given in Annex \ref{An:Setup}}. 

The dielectric measurements are performed by confining the polymer film between two disc-shaped electrodes of 10~cm diameter. 
The good contact between the electrodes and the sample, during the whole  experiment, is assured by  an aqueous gel, 
which  has very low electrical resistivity compared to the sample. We checked that the gel does not perturb the sample response because water absorption in our samples  is only 0.2\% 
{ and by comparing the response in the presence of the aqueous gel and in the presence of mineral oil \cite{perez2015simultaneous}}.

Dielectric spectroscopy allows the investigation of the dielectric response of a material as a function of the angular frequency
$\omega$.  It  is  expressed by the complex dielectric permittivity or dielectric constant: $\varepsilon(\omega)= \varepsilon'(\omega)- i\varepsilon''(\omega)$,
with $\varepsilon'(\omega)$ the real component of  the dielectric constant  which is related to the electric energy stored by the sample, 
and $\varepsilon''(\omega)$ the imaginary component which indicates the  energy losses. The loss tangent is
$\tnd(\omega)= \tan \delta(\omega)=\varepsilon''(\omega)/ \varepsilon'(\omega)$,
where $\delta$ is the  phase shift between the electric field applied for measuring $\varepsilon$ and the measured dielectric polarization.

We investigated an extruded film MAKROPOL\textsuperscript{\textregistered} DE 1-1 000000 (from BAYER) based on Makrolon\textsuperscript{\textregistered} polycarbonate (PC)  with a $T_g$ of about 150~$^\circ$C. The sample sheets had a thickness of 125~$\mu$m. 
The tensile experiments are performed at  $T=25^\circ$C much below $T_g$. We fixed four different strain rates:  $2.5\times10^{-3}$~s$^{-1}$, $2.5\times10^{-4}$~s$^{-1}$, $2.5\times10^{-5}$~s$^{-1}$ and $2.5\times10^{-6}$~s$^{-1}$.  
Typical strain curves at the different strain rates are plotted in  Fig.\ref{Fig:SetUp}.  
Initially, the stress increases almost linearly up to its maximum (yield stress) which is reached at $\lambda\simeq   1.08$. 
The stress decreases after the yield strain softening regime for  $1.09<\lambda <1.2$. For  $\lambda>1.2$, 
the material is inside the strain hardening regime, where the force increases up to the film crack.


\begin{figure}[h]
	\centering  
	\includegraphics[width=0.95\linewidth]{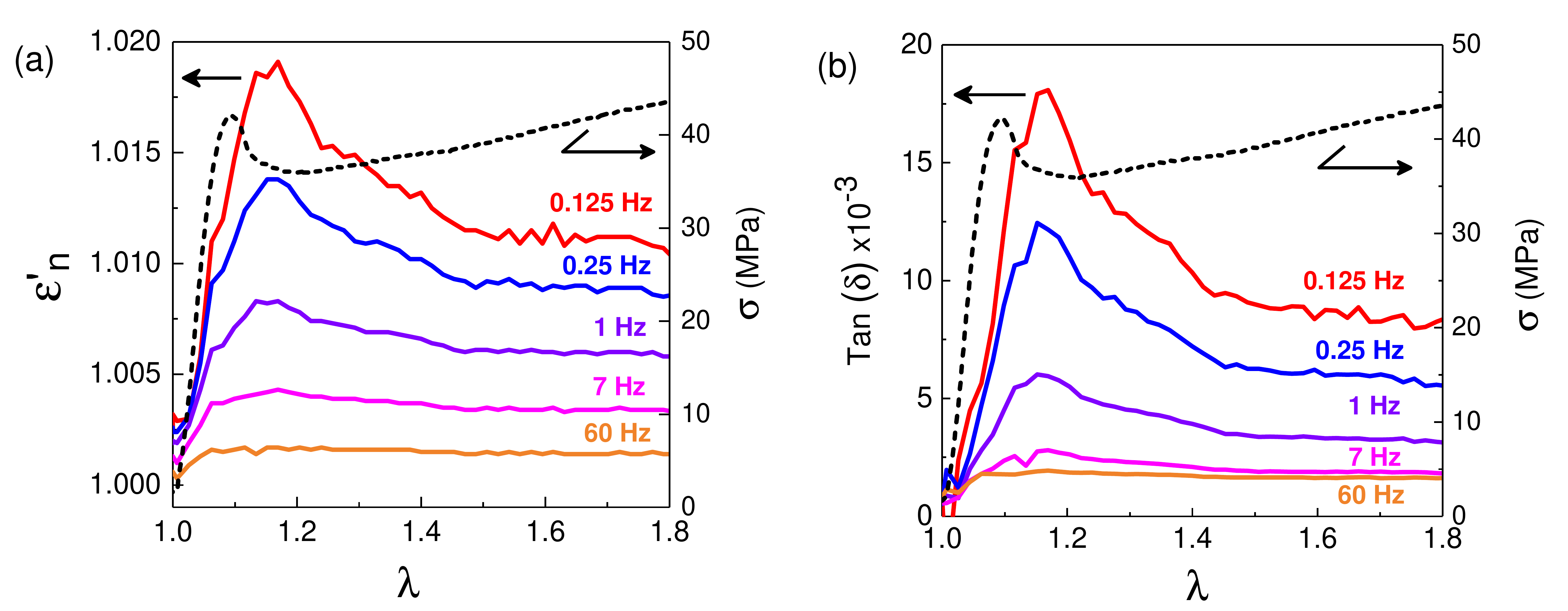}
	\caption{Dependence on $\lambda$ of $\epn$ (a) and $\tnd$ (b) measured  at several frequencies
		during the stretching experiment at $\dgam=2.5 \times 10^{-4} s^{-1}$.
		In both panels the solid lines (left hand ordinate) represent $\epn$(a) and $\tnd$(b) as a function of $\lambda$, whereas
		the dashed line (right hand ordinate) show  the corresponding evolution of the stress, which reveals 
		that the low frequency components of \DSs have a maximum in the softening regime of the polymer film.   
	}
	\label{Fig:EpsVsStrainVsFreq}
\end{figure}

\begin{figure}[h!]
  	\centering  
  	\includegraphics[width=0.9\linewidth]{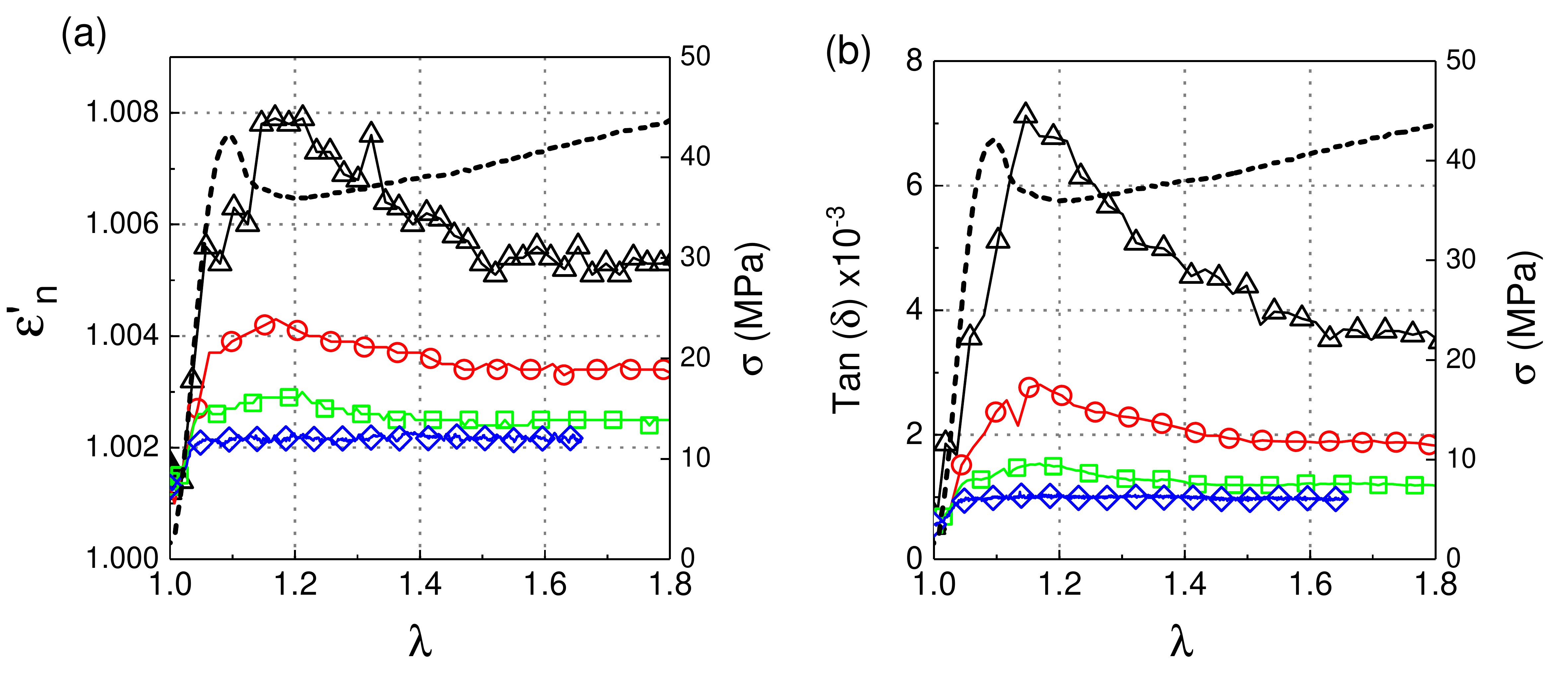}
  \caption{
  		Normalized real part $\epn$ (a) and $\tnd$ (b) as a function of $\lambda$ measured at $7$Hz for different strain rates : 
  		{$2.5\times10^{-3} s^{-1}$ (black triangles), $2.5\times10^{-4} s^{-1}$ (red circles), $2.5\times10^{-5} s^{-1}$ (green squares) and $2.5\times10^{-6} s^{-1}$ (blue diamonds). }
  	}
  	\label{Fig:EpsVsFreqVsStrainRate}
  \end{figure}

In this article we focus on the measurement  at $\lambda>1.09$ being the results 
at $\lambda \leq 1.08$ already discussed in~\cite{perez2016dielectric}. 
For each value of the strain rate the measurement of the dielectric and mechanical  properties  have  been repeated on
at least 3 samples, in order to check the reproducibility of the results. The maximum fluctuations of the results
 observed in  different samples is at most $11\%$.  In order to compensate for the change of the thickness of the sample we study  $\epn (f)=\varepsilon '(f,\lambda)/\varepsilon'(400Hz,\lambda)$ because above 400Hz the only effect of the applied stress on the dielectric measurement is related to the  change of the sample thickness induced by the large applied stress. 
 The quality of this compensation can be checked looking at Figs.\ref{Fig:EpsVsStrainVsFreq} where we plot $\epn(f,\lambda)$
 and $\tnd(f,\lambda)$ measured at various frequencies as  a function of $\lambda$.  We clearly see that  the effect of the stretching on the DS decreases a lot by increasing the measuring frequency and it almost disappears at $60$Hz. 
 This means that $\epn$ cancels the dependence { of the sample thickness} $d(\lambda)$ and correctly estimates the variation of $\varepsilon$ induced by the strain. This figure also shows that  both $\epn$ and $\tnd$ reach the maximum in the softening regime then  they decrease and remain constant in the hardening regimes. Fig.\ref{Fig:EpsVsFreqVsStrainRate} shows that the effect of the strain on $\varepsilon$ increases with $\dgam$ as already shown in ref.\cite{perez2016dielectric}.
 
  \begin{figure}[h!]
  	\centering  
  	\includegraphics[width=0.75\linewidth]{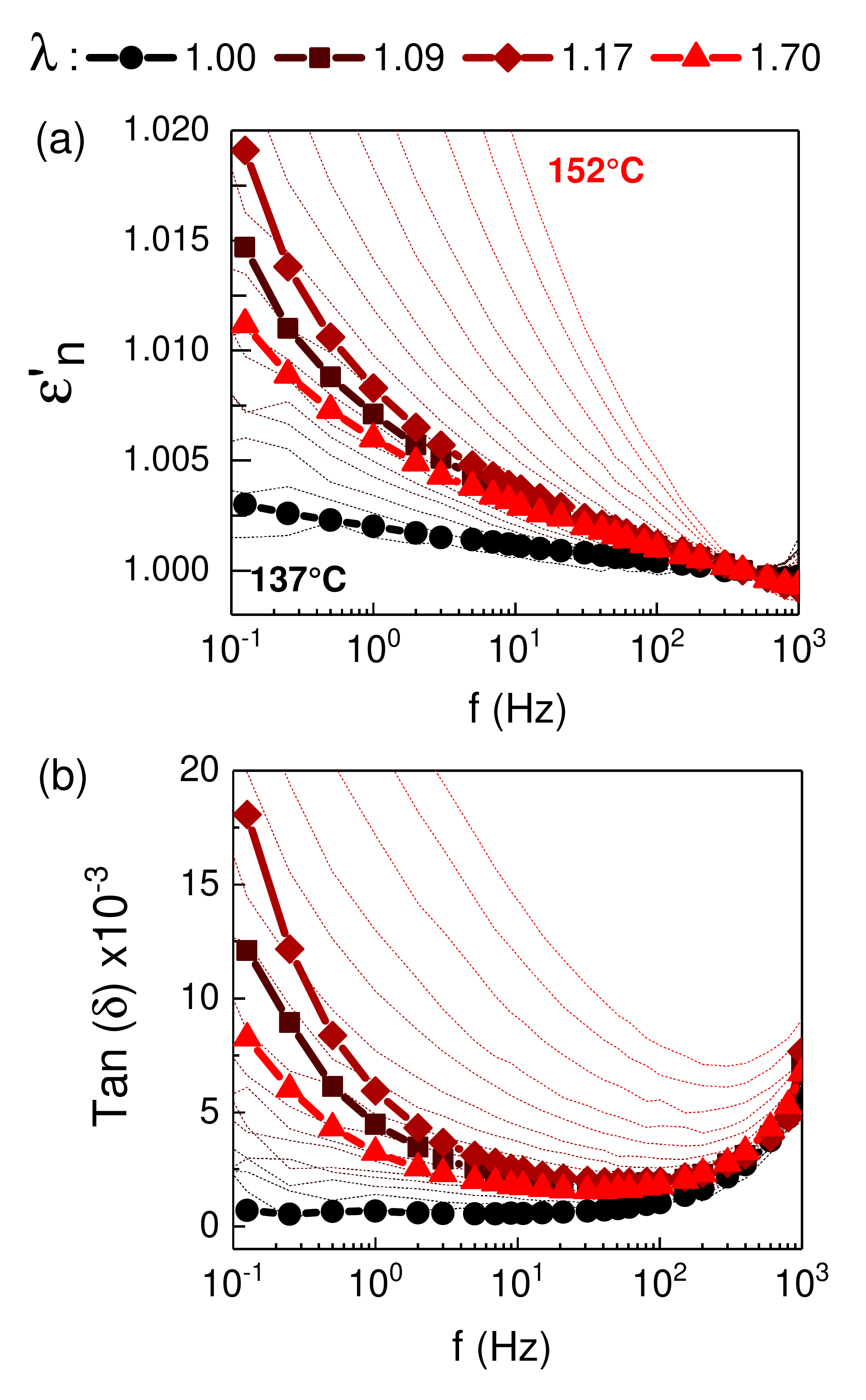}
  	\caption{
  		The normalized real part   $\epn$ (a) and $\tnd$ (b) as a function of frequency measured, at $T=25^\circ$C, during a  stretching experiment at $2.5\times10^{-4} s^{-1}$  at different values of the strain : $\lambda = 1.00$ (equilibrium), $\lambda = 1.08$ (plastic yield), $\lambda = 1.17$ 
  		(softening regime) and $\lambda = 1.70$  (hardening regime). In the background, continuous  orange lines represent  $\epn$ (a) and $\tnd$ (b) measured at $\lambda=1$ in  the temperature range    between  $137 ^\circ C$, (bottom orange curve)  and  { $152 ^\circ C$ (top orange  curve)} at  $1^\circ C$ increment  \cite{perez2016dielectric}.
  	}
  	\label{Fig:EpsVsFreq}
  \end{figure}

It is useful to study  the evolution of the whole \DSs measured  at $\dgam = 2.5\times10^{-4} s^{-1}$ for which  the effect of the strain on $\varepsilon$ is very pronounced (see Fig.\ref{Fig:EpsVsStrainVsFreq}) and at the same time the measure lasts enough time to have a good low frequency resolution  of the DS. 
 Figures~\ref{Fig:EpsVsFreq} show $\epn(f)$ and $\tnd(f)$ as a function of frequency measured  at various  $\lambda$  at $\dgam = 2.5\times10^{-4} s^{-1}$.  { The \DSs recorded  during the tensile test are compared to the \DST, i.e. the DS measured as a function of temperature in the unstressed sample.  This figure summarizes one the most important findings of this investigation, that we will explain. First we  notice that  the low frequency part of \DSs has roughly the same amplitude as \DST measured at  temperatures close to $T_g$. However the  $\tnd(f)$ of \DSs and of \DST have a very different  dependence on $f$ indicating that a simple relationship between $\dgam$ and an effective temperature ( presented  in several articles \cite{egami2010,chen2011,hebert2015,perez2016dielectric,perez2016dielectric}) cannot be easily  established as we discuss later on.}
 However in Figs.\ref{Fig:EpsVsFreq},\ref{Fig:EpsVsStrainVsFreq}  we notice several  other important results. We clearly see that in the stressed sample  the maximum amplitude of  $\epn(f)$  and $\tnd$ at low frequencies is not reached at the yield but in the softening regime at $\lambda=1.17$ and most importantly the amplitude of \DSs decreases  in the  hardening regime at $\lambda=1.70$.  This is a new and totally unexpected result, i.e.\ the effect of the strain on the dielectric constant is not monotonous. 
 \begin{figure}[ht]
	\centering  
	\includegraphics[width=0.95\linewidth]{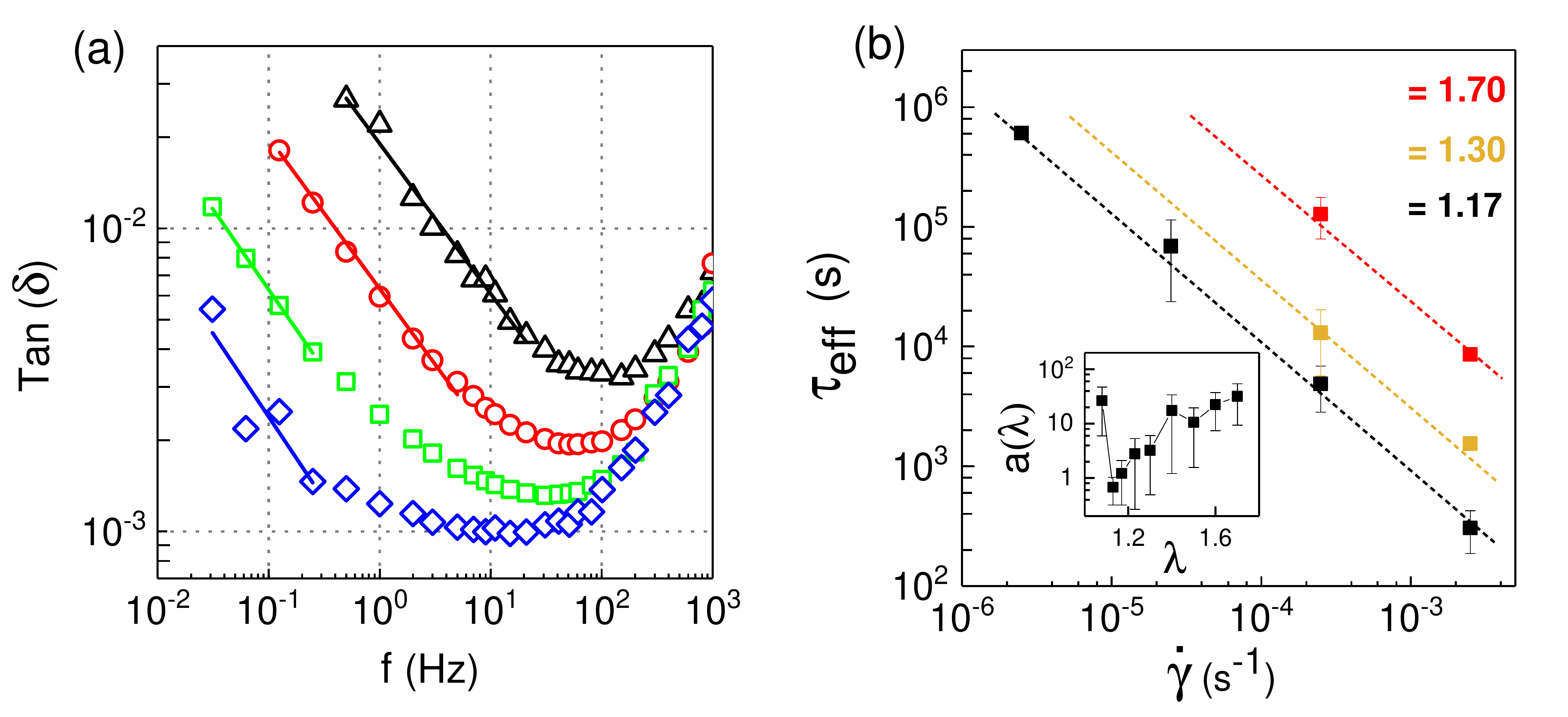}
	\caption{
		(a)	Dependence on frequency of $\tnd$, measured in the softening regime $\lambda = 1.17$, at different  
		$\dgam$: {$2.5\times10^{-3} s^{-1}$ (black triangles), $2.5\times10^{-4} s^{-1}$ (red circles), 
		$2.5\times 10^{-5} s^{-1}$ (green squares) and $2.5\times10^{-6} s^{-1}$ (blue diamonds). }The low frequency parts of each 
		curve are fitted by a  power law  (straight lines) $\tnd=1/(\ome  \tau_{eff})^\beta$
		The time scale  $\tau_{eff}$ depends on $\dgam$ but the scaling power {$\beta=0.50\pm 0.05$} is found to be constant within experimental errors.   (b) 	The fitted time scale $\tau_{eff}$ as a function of the strain rate $\dgam$ measured at several strains: $\lambda = 1.17$ (softening regime), $\lambda = 1.30$ and $\lambda = 1.70$ (hardening regime). The relaxation time follows the law 
		$\tau_{eff} = a(\lambda)/\dgam$ (dashed lines) with the coefficient $a(\lambda)$ plotted in the inset.
	}
	\label{Fig:FitTan}
\end{figure}

{ In order  to have a more detailed description of these experimental observations,  we study the behavior of $\tnd(f,\lambda,\dgam)$, which is not affected by any geometrical effect. }
 We plot, 
 in Fig.\ref{Fig:FitTan},  
 $\tnd(f)$
 versus frequency 
 measured at $\lambda=1.17$ at various $\dgam$.  We notice that the low frequency parts decrease as a power law
 which can be fitted with  { $\tnd (\omega )= 1/(\omega \tau_{eff}  ) ^{\beta} $} where  $\tau_{eff}$ and $\beta$ are fitting  parameters and $\omega=2 \pi f$. We find that
 { $\beta \simeq 0.50\pm 0.05$} is independent of $\dgam $ 
 (continuous lines in Fig.\ref{Fig:FitTan}a),{ with an accuracy of about $10\%$} . The measured  time scale  $\tau_{eff}$, which  is a function of $\dgam$ and $\lambda$, is  plotted in Fig.\ref{Fig:FitTan}b). Looking at this figure we see that with a very good accuracy { $\tau_{eff}= a(\lambda)/ \dgam$} where $a(\lambda )$ has a minimum  at $\lambda\simeq 1.17$ in the stress softening regime. 

{ We can consider the  power law  behavior of $\tnd$  as the high frequency part of the Cole-Cole (Havriliak-Negami) model \cite{havriliak1966} for the $\alpha$ peak of the dielectric constant. Indeed, calling $\tau$ the characteristic time of the $\alpha$ peak, at  frequencies $\ome>>1/\tau$ the  Cole-Cole model takes the form:} 
 \begin{equation}
 {\varepsilon'\over \epinf}-1\simeq {(\eps_o-\epinf)  \over \epinf }\ \ { \cos(\beta\pi/2) \over (\ome\tau)^{\beta}}
 \end{equation}
 and
 \begin{equation}
 \tnd \simeq{(\eps_o-\epinf) \over \epinf }\ \ { \sin(\beta\pi/2)    \over  (\ome\tau)^{\beta}}
 \label{eq:tand}
 \end{equation}
{ where  $\epinf$ and $\eps_o$ are respectively  the high and low  frequency dielectric constants.
	In our experiment $\beta\simeq 0.5$ thus the two equations become: 
}
 \begin{equation}
{\varepsilon'\over \epinf}-1\simeq {(\eps_o-\epinf)  \over \epinf }\ \ {  1 \over (2 \ome\tau)^{1/2}},
\end{equation}
 \begin{equation}
\tnd \simeq{ (\eps_o-\epinf) \over \epinf }\ \ { 1    \over  (2 \ome\tau)^{1/2}}.
\label{eq:tand_exp}
\end{equation}
{  These two equations show  that $\tnd \simeq ({\eps'\over \epinf}-1)$ which is rather well verified by the experimental data for all $\lambda$. This indicates  that they are a rather good model of the dielectric constant under stress. Using eq.\ref{eq:tand_exp}, we identify the fitting parameter as $\tau_{eff}=2\tau / \Delta^2$ where $\Delta={ (\eps_o-\epinf)/ \epinf }$. At this level we cannot distinguish whether the increase of $\eps'$ and $\tnd$ is induced by a decreasing of $\tau$ or an increase of the { relative dielectric strength} $\Delta$\cite{frohlich1958}.} 

	{ We suppose first that the stress has no influence on $\tau$ but only on $\Delta$.  Keeping for $\tau$ the value of the unstressed sample we  can estimate the value of $\Delta$ under stress from the previous expression  $\Delta=\sqrt{2\tau/\tau_{eff}}$. To give a lower bound on the value of $\tau$ at $22^oC$  we rely upon the high  temperature measurements of the dielectric properties of PC and on the estimation based on the  WLF law \cite{huang_new_2001,perez2016dielectric}. For example at $T=130^oC$, which is the minimum temperature at  which the WLF law for PC can be tested, one finds using WLF $\tau\simeq 10^7s$. Thus, as $\tau$ increases by lowering temperature, we simply assume that   at  $T=22^oC$, $\tau>10^7s$, which, in addition, is coherent with the fact that our samples are more than one year old \cite{Struik}. \\		
From the data of fig.\ref{Fig:FitTan}b) we obtain $\tau_{eff}\simeq 1/\dgam$ at $\lambda=1.17$. Thus in the hypothesis of constant $\tau$, we find  $\Delta>7$ at $\dgam=2.5 \times 10^{-6}$ and $\Delta>223$ at $\dgam=2.5 \times 10^{-3}s^{-1}$. This value is 1000 times larger than that  of the unstressed sample for which   $\Delta\simeq 0.12$. This is  nonphysical because  $\Delta$ cannot  change of such a large amount even at very large strain (see refs.\cite{frohlich1958,yee1981,rubber2013} and appendix \ref{An:Kirkwood}). Therefore one has to conclude that, although a small increase  of the $\Delta$  cannot be excluded, the observations cannot be explained without  a decreasing of $\tau$ with the mechanical stress, i.e. an  acceleration of the dynamics, which agrees  with other experimental results based on other techniques and other polymers \cite{ediger2000,lee2009,bending2014}. Using our data we can estimate $\tau  \simeq 0.5 \Delta^2 a(\lambda)/\dgam$, assuming that $\Delta=0.2$, i.e. keeping fix the value of the unstressed sample. At $\lambda=1.17$ we see that  $a(\lambda)\simeq 1$ and using $\dgam=2.5 \times 10^{-3}$ one finds $\tau= 8 s$, which implies that one would not observe a clear power law at the highest strain rate because the approximation $\ome>>1/\tau$ will be not fully satisfied  for the low frequencies of  our frequency range. Therefore at the highest $\dgam$, one concludes that $\tau>>8$ s and  $\Delta>0.2$. 
  At this point we can safely say that  the main effect of the strain  is a reduction of $\tau$ of several orders of magnitude  although a small influence on $\Delta$ is necessary for the consistency of the observations.}

{ Finally we compare  these results with those that one extracts from the values  $\tnd(f)$  measured  in the unstressed sample at different temperatures. In the range $137^oC <T<152^oC$, where we observe an overlap of \DSs and \DST  in fig.\ref{Fig:EpsVsStrainVsFreq}  the $\tnd(f)$ has  a power law dependence on $f$ but the  exponent $\beta$ is a function of temperature: specifically { $0.25<\beta<0.35$} for $137^oC <T<152^oC$.
	The observation that the $\beta$ exponent measured under stress is larger  than the one measured in the unstressed sample at temperature close to $T_g$, has important consequences. Indeed in  the Cole-Cole  expression   the smallest is  the value of the exponent $\beta$  the broadest is the distribution of relaxation times. 
	Thus we conclude that our measurements are compatible with a narrowing of the distribution of relaxation times under stress 
	observed in refs.\cite{lee2009,bending2014} with other techniques in other polymers.}	

{   Let us point out  that in the frequency window of our measurements there are no other relaxations which may influence the results. The beta relaxation is not affected by the stress because \DSs does not change above 400Hz. The gamma relaxation time is at  very high frequencies at room temperature (see ref\cite{yee1981})  }  
{   Structural changes that might also influence the results are absent. Indeed  we checked by Differential Scanning Calorimetry (DSC) that polycarbonate 
	does not crystallize under strain in the range considered here. Microscopic damaging may also contribute to the dielectric response. It has been shown that damaging occurs indeed during the strain hardening regime, but the volume fraction corresponding to these damages is very small ($< 10^{-4}$ for cellulose acetate  \cite{charvet2019} and for polycarbonate \cite{djukic2019}) and the corresponding perturbation regarding the interpretation of the results is negligible. }

Summarizing, we have studied the \DSs of a sample of polycarbonate at room temperature submitted to an applied stress at different strain rates covering three order of magnitude range. 
{ In our frequency window we observe that, with respect to the unstressed sample,  $\tnd$ increases at low frequencies as a function of $\dgam$, whereas at high frequencies it remains unchanged. The increase is not monotonous and reaches its maximum in the softening regime. From the data we extract an effective time scale $\tau_{eff}\propto 1/\dgam$ which depends on $\lambda$ and reaches its minimum in the softening regime to increase again in the hardening. We have also shown  that the distribution of relaxation times is narrower under stress than in the unstressed sample at $T$ close to $T_g$.} 
{ The  large variation of $\tau_{eff}$ cannot be explained only by the increase of the $\Delta$, but it implies  that the stress induces  a shift towards high frequencies of the $\alpha$ peak without  influencing  the high frequency part of the DS. 
Thus  we  confirm that the stress accelerates the polymer dynamics as already observed in other experiments \cite{loo2000,bending2014,kalfus2012,hebert2015,perez2016dielectric,ediger2000}, which were limited to the yield points. We have extended the analysis to high values of strain in a polymer which presents stress hardening, finding two very important and  unexpected results. Firstly the smallest $\tau_{eff}$  is not reached at yield but in the stress softening regime (see Figs.\ref{Fig:EpsVsStrainVsFreq},\ref{Fig:EpsVsFreqVsStrainRate}). Secondly the stress hardening regime  is associated to a progressive increase of the effective relaxation time $\tau_{eff}=2\tau / \Delta^2$.  This increase may  again be due either to an increase of $\tau$ or to a decrease of  $\Delta$  of at least  a factor of 3 (see  fig.\ref{Fig:FitTan}), which, on the basis of the previous arguments,  is  too large. Furthermore this decrease of $\Delta$ will imply a non monotonous dependence of  $\Delta$ as a function of $\lambda$, which contradicts previous measurements \cite{rubber2013,vogt1990}. Specifically  in ref. \cite{vogt1990} the authors measured by NMR a segmental 
orientation for stretched polycarbonate at room temperature, observing a monotonous increase of the orientational  
order parameter, whereas we observe a non monotonous behaviour of $\tau_{eff}$ versus $\lambda$. Thus it is conceivable to say that such a behavior is induced by an increase of $\tau$ in the hardening regime.}

As a conclusion our experimental results impose strong constrains on the theoretical models on strain softening and hardening. 

\section{Appendix}
\subsection{Experimental methods}\label{An:Setup}
The mechanical part is composed by two cylinders (21~cm long, 1.6 cm of diameter) used to fasten the sample, a load cell of 2000~N capacity, 
a precision linear transducer to measure the strain over a range of { 28 cm}, and a brushless servo motor with a coaxial reducer.
This device allows the investigation of polymer film samples of a maximum width of 21~cm. We use dog-bone shaped sheets  with the effective dimensions of the area under deformation of  {16X16~cm$^2$}, in order to focus the deformation of the film between the electrodes used to measure the DS. The sample is stretched  at fixed relative strain rate $\dgam=\frac{d}{dt}(L/L_o)$ 
Where $L(t)$ is  the length of the stretched  sample and  $L_o$  the initial length at zero applied force $F$. 
 Measurements are performed at room temperature at constant $\dgam$, till a   stretch ratio $\lambda = L(t)/L_0$ of about 1.8 (80 \% strain) is reached. 
The force $F(\lambda)$ applied to the sample  is measured during all the experiment by a load cell.  The value of the applied stress is defined  as $\sigma=F(\lambda)/(W_o \ d_o)$ where $W_o$ and $d_o$ are the initial width and thickness of the sample. 

 We use an innovative dielectric spectroscopy technique (see ref. \cite{perez2015simultaneous}), which allows the simultaneous measurement of the dielectric properties 
in a four orders of magnitude frequency window chosen in the range of $10^{-2}-10^3$~Hz.  
This multi-frequency experiment is very useful in the study of transient phenomena such as polymer films deformation (see~\cite{perez2015simultaneous} for more details),  because it gives the evolution of the DS on a wide frequency range instead of a single frequency. Specifically the device measures the complex  impedance of the capacitance $C$ formed by the electrodes and the sample. In all the frequency range  the device has an  accuracy better than $1\%$ on the measure of $C$   and  it  can detect  values of $\tnd $ smaller than $10^{-4}$.  The real part of the dielectric constant of the sample is $\varepsilon' =C d(\lambda)/(S\varepsilon_o)$ where $S$ is the electrode surface, $\varepsilon_o$ the vacuum dielectric constant and $d(\lambda)$ the thickness of the sample whose dependence on $\lambda$ must be taken into account  to correctly estimate the value of $\varepsilon'$.   This compensation is not necessary for $\tnd (f)$ 
because it does not depend on the geometry being the ratio of $\varepsilon''/\varepsilon'$.

\subsection{Kirkwood factor}\label{An:Kirkwood}

{ The low frequency dielectric constant $\eps_o$ is related to $\epsilon_\infty$ via the Kirkwood factor  $g = 1 + z <\cos\theta_{i,j}>$ where $<.>$ stands for mean value,  $\theta_{i,j}$ is 
	the angle between dipole i and dipole j, and  $z$ is the number of relevant nearest dipoles, which is usually taken to be about $10 \ $ (see ref. \cite{frohlich1958}). Specifically one finds that $\eps_o=\eps_\infty+ A \ g$ where $A$ is a material dependent factor. Therefore  using this relationship, the  quantity  $\Delta$  defined in the text  becomes:  $\Delta= \eps_o/\eps_\infty-1 = A g/\eps_\infty$. 
	
 We notice that, independently of the probability distribution  of  $\theta_{i,j}$, the value of $ <\cos\theta_{i,j}> $ is limited between $0$ and $1$ and  as a consequence $1<g<11$.  This means that, being $\Delta=0.2$ in the unstressed sample, the maximum value of $\Delta$ under stress is $\Delta=2.2$. This value is much smaller than the values of $\Delta$   estimated in the main text in order to explain the observations, assuming that only $\Delta$ is affected by the strain.  Thus  on the basis of the experimental data,  we have to conclude that $\tau$ decreases in the stressed sample, i.e.   the dynamics accelerates  under strain. However this acceleration is associated to a small increase of $\Delta$ as we discuss in the main text. }

\bigskip
$^*$ Present address of Riad Sahli: INM-Leibniz Institute for New Materials, 66123, Saarbr$\ddot{\textrm{u}}$cken, Germany

\bibliographystyle{apsrev4-1}
\bibliography{authors}

\begin{thebibliography}{41}%
\makeatletter
\providecommand \@ifxundefined [1]{%
 \@ifx{#1\undefined}
}%
\providecommand \@ifnum [1]{%
 \ifnum #1\expandafter \@firstoftwo
 \else \expandafter \@secondoftwo
 \fi
}%
\providecommand \@ifx [1]{%
 \ifx #1\expandafter \@firstoftwo
 \else \expandafter \@secondoftwo
 \fi
}%
\providecommand \natexlab [1]{#1}%
\providecommand \enquote  [1]{``#1''}%
\providecommand \bibnamefont  [1]{#1}%
\providecommand \bibfnamefont [1]{#1}%
\providecommand \citenamefont [1]{#1}%
\providecommand \href@noop [0]{\@secondoftwo}%
\providecommand \href [0]{\begingroup \@sanitize@url \@href}%
\providecommand \@href[1]{\@@startlink{#1}\@@href}%
\providecommand \@@href[1]{\endgroup#1\@@endlink}%
\providecommand \@sanitize@url [0]{\catcode `\\12\catcode `\$12\catcode
  `\&12\catcode `\#12\catcode `\^12\catcode `\_12\catcode `\%12\relax}%
\providecommand \@@startlink[1]{}%
\providecommand \@@endlink[0]{}%
\providecommand \url  [0]{\begingroup\@sanitize@url \@url }%
\providecommand \@url [1]{\endgroup\@href {#1}{\urlprefix }}%
\providecommand \urlprefix  [0]{URL }%
\providecommand \Eprint [0]{\href }%
\providecommand \doibase [0]{http://dx.doi.org/}%
\providecommand \selectlanguage [0]{\@gobble}%
\providecommand \bibinfo  [0]{\@secondoftwo}%
\providecommand \bibfield  [0]{\@secondoftwo}%
\providecommand \translation [1]{[#1]}%
\providecommand \BibitemOpen [0]{}%
\providecommand \bibitemStop [0]{}%
\providecommand \bibitemNoStop [0]{.\EOS\space}%
\providecommand \EOS [0]{\spacefactor3000\relax}%
\providecommand \BibitemShut  [1]{\csname bibitem#1\endcsname}%
\let\auto@bib@innerbib\@empty
\bibitem [{\citenamefont {Ferry}(1980)}]{Ferry}%
  \BibitemOpen
  \bibfield  {author} {\bibinfo {author} {\bibfnamefont {J.~D.}\ \bibnamefont
  {Ferry}},\ }\href@noop {} {\emph {\bibinfo {title} {Viscoelastic Properties
  of Polymers}}}\ (\bibinfo  {publisher} {John Wiley and Sons, Inc.},\ \bibinfo
  {year} {1980})\BibitemShut {NoStop}%
\bibitem [{\citenamefont {Nielsen}\ and\ \citenamefont
  {Landel}(1994)}]{landel}%
  \BibitemOpen
  \bibfield  {author} {\bibinfo {author} {\bibfnamefont {L.~E.}\ \bibnamefont
  {Nielsen}}\ and\ \bibinfo {author} {\bibfnamefont {R.~F.}\ \bibnamefont
  {Landel}},\ }\href@noop {} {\emph {\bibinfo {title} {Mechanical Properties of
  Polymers and Composites}}}\ (\bibinfo  {publisher} {Marcel Dekker, New
  York},\ \bibinfo {year} {1994})\BibitemShut {NoStop}%
\bibitem [{\citenamefont {Meijer}\ and\ \citenamefont
  {Govaert}(2003)}]{Meijer2003}%
  \BibitemOpen
  \bibfield  {author} {\bibinfo {author} {\bibfnamefont {H.~E.~H.}\
  \bibnamefont {Meijer}}\ and\ \bibinfo {author} {\bibfnamefont {L.~E.}\
  \bibnamefont {Govaert}},\ }\href@noop {} {\bibfield  {journal} {\bibinfo
  {journal} {Macromolecular Chemistry and Physics}\ }\textbf {\bibinfo {volume}
  {204}},\ \bibinfo {pages} {274} (\bibinfo {year} {2003})}\BibitemShut
  {NoStop}%
\bibitem [{\citenamefont {Meijer}\ and\ \citenamefont
  {Govaert}(2005)}]{meijer2005}%
  \BibitemOpen
  \bibfield  {author} {\bibinfo {author} {\bibfnamefont {H.~E.~H.}\
  \bibnamefont {Meijer}}\ and\ \bibinfo {author} {\bibfnamefont {L.~E.}\
  \bibnamefont {Govaert}},\ }\href@noop {} {\bibfield  {journal} {\bibinfo
  {journal} {Progress in Polymer Science}\ }\textbf {\bibinfo {volume} {30}},\
  \bibinfo {pages} {915} (\bibinfo {year} {2005})}\BibitemShut {NoStop}%
\bibitem [{\citenamefont {Hoy}\ and\ \citenamefont {Robbins}(2006)}]{hoy2006b}%
  \BibitemOpen
  \bibfield  {author} {\bibinfo {author} {\bibfnamefont {R.~S.}\ \bibnamefont
  {Hoy}}\ and\ \bibinfo {author} {\bibfnamefont {M.~O.}\ \bibnamefont
  {Robbins}},\ }\href@noop {} {\bibfield  {journal} {\bibinfo  {journal} {The
  American Physical Society Division of polymer Physics}\ }\textbf {\bibinfo
  {volume} {44}},\ \bibinfo {pages} {3487} (\bibinfo {year}
  {2006})}\BibitemShut {NoStop}%
\bibitem [{\citenamefont {Hoy}\ and\ \citenamefont {O'Hern}(2010)}]{hoy2010}%
  \BibitemOpen
  \bibfield  {author} {\bibinfo {author} {\bibfnamefont {R.~S.}\ \bibnamefont
  {Hoy}}\ and\ \bibinfo {author} {\bibfnamefont {C.~S.}\ \bibnamefont
  {O'Hern}},\ }\href@noop {} {\bibfield  {journal} {\bibinfo  {journal}
  {Physical Review E}\ }\textbf {\bibinfo {volume} {82}} (\bibinfo {year}
  {2010})}\BibitemShut {NoStop}%
\bibitem [{\citenamefont {Hoy}(2011)}]{hoy2011}%
  \BibitemOpen
  \bibfield  {author} {\bibinfo {author} {\bibfnamefont {R.~S.}\ \bibnamefont
  {Hoy}},\ }\href@noop {} {\bibfield  {journal} {\bibinfo  {journal} {Journal
  of polymer science. Part B. Polymer physics}\ }\textbf {\bibinfo {volume}
  {49}},\ \bibinfo {pages} {979} (\bibinfo {year} {2011})}\BibitemShut
  {NoStop}%
\bibitem [{\citenamefont {Van~Melick}\ \emph {et~al.}(2003)\citenamefont
  {Van~Melick}, \citenamefont {Meijer},\ and\ \citenamefont
  {Govaert}}]{vanmelick2003}%
  \BibitemOpen
  \bibfield  {author} {\bibinfo {author} {\bibfnamefont {H.~G.~H.}\
  \bibnamefont {Van~Melick}}, \bibinfo {author} {\bibfnamefont {H.~E.~H.}\
  \bibnamefont {Meijer}}, \ and\ \bibinfo {author} {\bibfnamefont {L.~E.}\
  \bibnamefont {Govaert}},\ }\href@noop {} {\bibfield  {journal} {\bibinfo
  {journal} {Polymer}\ }\textbf {\bibinfo {volume} {44}},\ \bibinfo {pages}
  {2493} (\bibinfo {year} {2003})}\BibitemShut {NoStop}%
\bibitem [{\citenamefont {Govaert}\ \emph {et~al.}(2008)\citenamefont
  {Govaert}, \citenamefont {Engels}, \citenamefont {Wendlandt}, \citenamefont
  {Tervoort},\ and\ \citenamefont {Suter}}]{govaert2008}%
  \BibitemOpen
  \bibfield  {author} {\bibinfo {author} {\bibfnamefont {L.~E.}\ \bibnamefont
  {Govaert}}, \bibinfo {author} {\bibfnamefont {T.~A.~P.}\ \bibnamefont
  {Engels}}, \bibinfo {author} {\bibfnamefont {M.}~\bibnamefont {Wendlandt}},
  \bibinfo {author} {\bibfnamefont {T.~A.}\ \bibnamefont {Tervoort}}, \ and\
  \bibinfo {author} {\bibfnamefont {U.~W.}\ \bibnamefont {Suter}},\ }\href@noop
  {} {\bibfield  {journal} {\bibinfo  {journal} {Journal of polymer science.
  Part B. Polymer physics}\ }\textbf {\bibinfo {volume} {46}},\ \bibinfo
  {pages} {2475} (\bibinfo {year} {2008})}\BibitemShut {NoStop}%
\bibitem [{\citenamefont {Senden}\ \emph {et~al.}(2010)\citenamefont {Senden},
  \citenamefont {van Dommelen},\ and\ \citenamefont {Govaert}}]{senden2010}%
  \BibitemOpen
  \bibfield  {author} {\bibinfo {author} {\bibfnamefont {D.~J.~A.}\
  \bibnamefont {Senden}}, \bibinfo {author} {\bibfnamefont {J.~A.~W.}\
  \bibnamefont {van Dommelen}}, \ and\ \bibinfo {author} {\bibfnamefont
  {L.~E.}\ \bibnamefont {Govaert}},\ }\href@noop {} {\bibfield  {journal}
  {\bibinfo  {journal} {Journal of Polymer Science Part B: Polymer Physics}\
  }\textbf {\bibinfo {volume} {48}},\ \bibinfo {pages} {1483} (\bibinfo {year}
  {2010})}\BibitemShut {NoStop}%
\bibitem [{\citenamefont {Jatin}\ \emph {et~al.}(2014)\citenamefont {Jatin},
  \citenamefont {Sudarkodi},\ and\ \citenamefont {Basu}}]{jatin2014}%
  \BibitemOpen
  \bibfield  {author} {\bibinfo {author} {\bibnamefont {Jatin}}, \bibinfo
  {author} {\bibfnamefont {V.}~\bibnamefont {Sudarkodi}}, \ and\ \bibinfo
  {author} {\bibfnamefont {S.}~\bibnamefont {Basu}},\ }\href@noop {} {\bibfield
   {journal} {\bibinfo  {journal} {International Journal of Plasticity}\
  }\textbf {\bibinfo {volume} {56}},\ \bibinfo {pages} {139} (\bibinfo {year}
  {2014})}\BibitemShut {NoStop}%
\bibitem [{\citenamefont {Chui}\ and\ \citenamefont {Boyce}(1999)}]{chui1999}%
  \BibitemOpen
  \bibfield  {author} {\bibinfo {author} {\bibfnamefont {C.}~\bibnamefont
  {Chui}}\ and\ \bibinfo {author} {\bibfnamefont {M.~C.}\ \bibnamefont
  {Boyce}},\ }\href@noop {} {\bibfield  {journal} {\bibinfo  {journal}
  {Macromolecules}\ }\textbf {\bibinfo {volume} {32}},\ \bibinfo {pages} {3795}
  (\bibinfo {year} {1999})}\BibitemShut {NoStop}%
\bibitem [{\citenamefont {Hasan}\ and\ \citenamefont
  {Boyce}(1993)}]{hasan1993}%
  \BibitemOpen
  \bibfield  {author} {\bibinfo {author} {\bibfnamefont {O.~A.}\ \bibnamefont
  {Hasan}}\ and\ \bibinfo {author} {\bibfnamefont {M.~C.}\ \bibnamefont
  {Boyce}},\ }\href@noop {} {\bibfield  {journal} {\bibinfo  {journal}
  {Polymer}\ }\textbf {\bibinfo {volume} {34}},\ \bibinfo {pages} {5085}
  (\bibinfo {year} {1993})}\BibitemShut {NoStop}%
\bibitem [{\citenamefont {Hoy}\ and\ \citenamefont
  {Robbins}(2009)}]{robbins2009}%
  \BibitemOpen
  \bibfield  {author} {\bibinfo {author} {\bibfnamefont {R.~S.}\ \bibnamefont
  {Hoy}}\ and\ \bibinfo {author} {\bibfnamefont {M.~O.}\ \bibnamefont
  {Robbins}},\ }\href@noop {} {\bibfield  {journal} {\bibinfo  {journal}
  {Polymer Physics}\ }\textbf {\bibinfo {volume} {47}},\ \bibinfo {pages}
  {1406} (\bibinfo {year} {2009})}\BibitemShut {NoStop}%
\bibitem [{\citenamefont {Ge}\ and\ \citenamefont {Robbins}(2010)}]{ge2010}%
  \BibitemOpen
  \bibfield  {author} {\bibinfo {author} {\bibfnamefont {T.}~\bibnamefont
  {Ge}}\ and\ \bibinfo {author} {\bibfnamefont {M.~O.}\ \bibnamefont
  {Robbins}},\ }\href@noop {} {\bibfield  {journal} {\bibinfo  {journal}
  {Journal of Polymer Science Part B: Polymer Physics}\ }\textbf {\bibinfo
  {volume} {48}},\ \bibinfo {pages} {1473} (\bibinfo {year}
  {2010})}\BibitemShut {NoStop}%
\bibitem [{\citenamefont {Kramer}(2005)}]{kramer2005}%
  \BibitemOpen
  \bibfield  {author} {\bibinfo {author} {\bibfnamefont {E.~J.}\ \bibnamefont
  {Kramer}},\ }\href@noop {} {\bibfield  {journal} {\bibinfo  {journal}
  {Journal of polymer science. Part B. Polymer physics}\ }\textbf {\bibinfo
  {volume} {43}},\ \bibinfo {pages} {3369} (\bibinfo {year}
  {2005})}\BibitemShut {NoStop}%
\bibitem [{\citenamefont {Hoy}(2016)}]{hoy2016}%
  \BibitemOpen
  \bibfield  {author} {\bibinfo {author} {\bibfnamefont {R.~S.}\ \bibnamefont
  {Hoy}},\ }\href@noop {} {\emph {\bibinfo {title} {Polymer Glasses. Chapter:
  Modeling strain hardening in polymer glasses using molecular simulations. pp
  425-450}}}\ (\bibinfo  {publisher} {Taylor and Francis Group: Boca Raton,
  Edited by Connie B. Roth},\ \bibinfo {year} {2016})\BibitemShut {NoStop}%
\bibitem [{\citenamefont {Nicolas}\ \emph {et~al.}(2018)\citenamefont
  {Nicolas}, \citenamefont {Ferrer}, \citenamefont {Martens},\ and\
  \citenamefont {Barrat}}]{rev_mod_phys_2018}%
  \BibitemOpen
  \bibfield  {author} {\bibinfo {author} {\bibfnamefont {A.}~\bibnamefont
  {Nicolas}}, \bibinfo {author} {\bibfnamefont {E.}~\bibnamefont {Ferrer}},
  \bibinfo {author} {\bibfnamefont {K.}~\bibnamefont {Martens}}, \ and\
  \bibinfo {author} {\bibfnamefont {J.~L.}\ \bibnamefont {Barrat}},\
  }\href@noop {} {\bibfield  {journal} {\bibinfo  {journal} {Rev. Mod. Phys.}\
  ,\ \bibinfo {pages} {045006}} (\bibinfo {year} {2018})}\BibitemShut {NoStop}%
\bibitem [{\citenamefont {Conca}\ \emph {et~al.}(2017)\citenamefont {Conca},
  \citenamefont {Dequidt}, \citenamefont {Sotta},\ and\ \citenamefont
  {Long}}]{conca2017}%
  \BibitemOpen
  \bibfield  {author} {\bibinfo {author} {\bibfnamefont {L.}~\bibnamefont
  {Conca}}, \bibinfo {author} {\bibfnamefont {A.}~\bibnamefont {Dequidt}},
  \bibinfo {author} {\bibfnamefont {P.}~\bibnamefont {Sotta}}, \ and\ \bibinfo
  {author} {\bibfnamefont {D.~R.}\ \bibnamefont {Long}},\ }\href@noop {}
  {\bibfield  {journal} {\bibinfo  {journal} {Macromolecules}\ }\textbf
  {\bibinfo {volume} {50}},\ \bibinfo {pages} {9456} (\bibinfo {year}
  {2017})}\BibitemShut {NoStop}%
\bibitem [{\citenamefont {Jaiswal}\ \emph {et~al.}(2016)\citenamefont
  {Jaiswal}, \citenamefont {Procaccia}, \citenamefont {Rainone},\ and\
  \citenamefont {Singh}}]{Procaccia2016}%
  \BibitemOpen
  \bibfield  {author} {\bibinfo {author} {\bibfnamefont {P.~K.}\ \bibnamefont
  {Jaiswal}}, \bibinfo {author} {\bibfnamefont {I.}~\bibnamefont {Procaccia}},
  \bibinfo {author} {\bibfnamefont {C.}~\bibnamefont {Rainone}}, \ and\
  \bibinfo {author} {\bibfnamefont {M.}~\bibnamefont {Singh}},\ }\href@noop {}
  {\bibfield  {journal} {\bibinfo  {journal} {Phys. Rev. Lett.}\ ,\ \bibinfo
  {pages} {085501}} (\bibinfo {year} {2016})}\BibitemShut {NoStop}%
\bibitem [{\citenamefont {Roth}(2016)}]{roth2016}%
  \BibitemOpen
  \bibfield  {author} {\bibinfo {author} {\bibfnamefont {C.~B.}\ \bibnamefont
  {Roth}},\ }\href@noop {} {\emph {\bibinfo {title} {Polymer Glasses}}}\
  (\bibinfo  {publisher} {Taylor and Francis Group: Boca Raton},\ \bibinfo
  {year} {2016})\BibitemShut {NoStop}%
\bibitem [{\citenamefont {Venkatasway}\ \emph {et~al.}(1982)\citenamefont
  {Venkatasway}, \citenamefont {Ard},\ and\ \citenamefont
  {Beatty}}]{venkataswamy1982}%
  \BibitemOpen
  \bibfield  {author} {\bibinfo {author} {\bibfnamefont {K.}~\bibnamefont
  {Venkatasway}}, \bibinfo {author} {\bibfnamefont {K.}~\bibnamefont {Ard}}, \
  and\ \bibinfo {author} {\bibfnamefont {C.~L.}\ \bibnamefont {Beatty}},\
  }\href@noop {} {\bibfield  {journal} {\bibinfo  {journal} {Polymer
  Engineering and Science}\ }\textbf {\bibinfo {volume} {22}},\ \bibinfo
  {pages} {961} (\bibinfo {year} {1982})}\BibitemShut {NoStop}%
\bibitem [{\citenamefont {Loo}\ \emph {et~al.}(2000)\citenamefont {Loo},
  \citenamefont {Cohen},\ and\ \citenamefont {Gleason}}]{loo2000}%
  \BibitemOpen
  \bibfield  {author} {\bibinfo {author} {\bibfnamefont {L.~S.}\ \bibnamefont
  {Loo}}, \bibinfo {author} {\bibfnamefont {R.~E.}\ \bibnamefont {Cohen}}, \
  and\ \bibinfo {author} {\bibfnamefont {K.~K.}\ \bibnamefont {Gleason}},\
  }\href@noop {} {\ \textbf {\bibinfo {volume} {288}},\ \bibinfo {pages}
  {116119} (\bibinfo {year} {2000})}\BibitemShut {NoStop}%
\bibitem [{\citenamefont {Ediger}(2000)}]{ediger2000}%
  \BibitemOpen
  \bibfield  {author} {\bibinfo {author} {\bibfnamefont {M.~D.}\ \bibnamefont
  {Ediger}},\ }\href@noop {} {\bibfield  {journal} {\bibinfo  {journal} {Annu.
  Rev. Chem.}\ }\textbf {\bibinfo {volume} {51}},\ \bibinfo {pages} {99}
  (\bibinfo {year} {2000})}\BibitemShut {NoStop}%
\bibitem [{\citenamefont {Bending}\ \emph {et~al.}(2014)\citenamefont
  {Bending}, \citenamefont {Christison}, \citenamefont {Ricci},\ and\
  \citenamefont {Ediger}}]{bending2014}%
  \BibitemOpen
  \bibfield  {author} {\bibinfo {author} {\bibfnamefont {B.}~\bibnamefont
  {Bending}}, \bibinfo {author} {\bibfnamefont {K.}~\bibnamefont {Christison}},
  \bibinfo {author} {\bibfnamefont {J.}~\bibnamefont {Ricci}}, \ and\ \bibinfo
  {author} {\bibfnamefont {M.~D.}\ \bibnamefont {Ediger}},\ }\href@noop {}
  {\bibfield  {journal} {\bibinfo  {journal} {Macromolecules}\ }\textbf
  {\bibinfo {volume} {47}},\ \bibinfo {pages} {800 } (\bibinfo {year}
  {2014})}\BibitemShut {NoStop}%
\bibitem [{\citenamefont {Hebert}\ \emph {et~al.}(2015)\citenamefont {Hebert},
  \citenamefont {Bending}, \citenamefont {Ricci},\ and\ \citenamefont
  {Ediger}}]{hebert2015}%
  \BibitemOpen
  \bibfield  {author} {\bibinfo {author} {\bibfnamefont {K.}~\bibnamefont
  {Hebert}}, \bibinfo {author} {\bibfnamefont {B.}~\bibnamefont {Bending}},
  \bibinfo {author} {\bibfnamefont {J.}~\bibnamefont {Ricci}}, \ and\ \bibinfo
  {author} {\bibfnamefont {M.~D.}\ \bibnamefont {Ediger}},\ }\href@noop {}
  {\bibfield  {journal} {\bibinfo  {journal} {Macromolecules}\ }\textbf
  {\bibinfo {volume} {48}},\ \bibinfo {pages} {6736} (\bibinfo {year}
  {2015})}\BibitemShut {NoStop}%
\bibitem [{\citenamefont {Kalfus}\ \emph {et~al.}(2012)\citenamefont {Kalfus},
  \citenamefont {Detwiler},\ and\ \citenamefont {Lesser}}]{kalfus2012}%
  \BibitemOpen
  \bibfield  {author} {\bibinfo {author} {\bibfnamefont {J.}~\bibnamefont
  {Kalfus}}, \bibinfo {author} {\bibfnamefont {A.}~\bibnamefont {Detwiler}}, \
  and\ \bibinfo {author} {\bibfnamefont {A.~J.}\ \bibnamefont {Lesser}},\
  }\href@noop {} {\bibfield  {journal} {\bibinfo  {journal} {Macromolecules}\
  }\textbf {\bibinfo {volume} {45}},\ \bibinfo {pages} {4839 } (\bibinfo {year}
  {2012})}\BibitemShut {NoStop}%
\bibitem [{\citenamefont {Perez-Aparicio}\ \emph {et~al.}(2016)\citenamefont
  {Perez-Aparicio}, \citenamefont {Cottinet}, \citenamefont
  {Crauste-Thibierge}, \citenamefont {Vanel}, \citenamefont {Sotta},
  \citenamefont {Delannoy}, \citenamefont {Long},\ and\ \citenamefont
  {Ciliberto}}]{perez2016dielectric}%
  \BibitemOpen
  \bibfield  {author} {\bibinfo {author} {\bibfnamefont {R.}~\bibnamefont
  {Perez-Aparicio}}, \bibinfo {author} {\bibfnamefont {D.}~\bibnamefont
  {Cottinet}}, \bibinfo {author} {\bibfnamefont {C.}~\bibnamefont
  {Crauste-Thibierge}}, \bibinfo {author} {\bibfnamefont {L.}~\bibnamefont
  {Vanel}}, \bibinfo {author} {\bibfnamefont {P.}~\bibnamefont {Sotta}},
  \bibinfo {author} {\bibfnamefont {J.-Y.}\ \bibnamefont {Delannoy}}, \bibinfo
  {author} {\bibfnamefont {D.~R.}\ \bibnamefont {Long}}, \ and\ \bibinfo
  {author} {\bibfnamefont {S.}~\bibnamefont {Ciliberto}},\ }\href@noop {}
  {\bibfield  {journal} {\bibinfo  {journal} {Macromolecules}\ }\textbf
  {\bibinfo {volume} {49}},\ \bibinfo {pages} {3889} (\bibinfo {year}
  {2016})}\BibitemShut {NoStop}%
\bibitem [{\citenamefont {P{\'e}rez-Aparicio}\ \emph
  {et~al.}(2015)\citenamefont {P{\'e}rez-Aparicio}, \citenamefont
  {Crauste-Thibierge}, \citenamefont {Cottinet}, \citenamefont {Tanase},
  \citenamefont {Metz}, \citenamefont {Bellon}, \citenamefont {Naert},\ and\
  \citenamefont {Ciliberto}}]{perez2015simultaneous}%
  \BibitemOpen
  \bibfield  {author} {\bibinfo {author} {\bibfnamefont {R.}~\bibnamefont
  {P{\'e}rez-Aparicio}}, \bibinfo {author} {\bibfnamefont {C.}~\bibnamefont
  {Crauste-Thibierge}}, \bibinfo {author} {\bibfnamefont {D.}~\bibnamefont
  {Cottinet}}, \bibinfo {author} {\bibfnamefont {M.}~\bibnamefont {Tanase}},
  \bibinfo {author} {\bibfnamefont {P.}~\bibnamefont {Metz}}, \bibinfo {author}
  {\bibfnamefont {L.}~\bibnamefont {Bellon}}, \bibinfo {author} {\bibfnamefont
  {A.}~\bibnamefont {Naert}}, \ and\ \bibinfo {author} {\bibfnamefont
  {S.}~\bibnamefont {Ciliberto}},\ }\href@noop {} {\bibfield  {journal}
  {\bibinfo  {journal} {Review of Scientific Instruments}\ }\textbf {\bibinfo
  {volume} {86}},\ \bibinfo {pages} {044702} (\bibinfo {year}
  {2015})}\BibitemShut {NoStop}%
\bibitem [{\citenamefont {Guan}\ \emph {et~al.}(2010)\citenamefont {Guan},
  \citenamefont {Chen},\ and\ \citenamefont {Egami}}]{egami2010}%
  \BibitemOpen
  \bibfield  {author} {\bibinfo {author} {\bibfnamefont {P.}~\bibnamefont
  {Guan}}, \bibinfo {author} {\bibfnamefont {M.}~\bibnamefont {Chen}}, \ and\
  \bibinfo {author} {\bibfnamefont {T.}~\bibnamefont {Egami}},\ }\href@noop {}
  {\bibfield  {journal} {\bibinfo  {journal} {Phys. Rev. Lett.}\ }\textbf
  {\bibinfo {volume} {104}},\ \bibinfo {pages} {205701} (\bibinfo {year}
  {2010})}\BibitemShut {NoStop}%
\bibitem [{\citenamefont {Chen}\ and\ \citenamefont
  {Schweizer}(2011)}]{chen2011}%
  \BibitemOpen
  \bibfield  {author} {\bibinfo {author} {\bibfnamefont {K.}~\bibnamefont
  {Chen}}\ and\ \bibinfo {author} {\bibfnamefont {K.~S.}\ \bibnamefont
  {Schweizer}},\ }\href@noop {} {\bibfield  {journal} {\bibinfo  {journal}
  {Macromolecules}\ }\textbf {\bibinfo {volume} {44}},\ \bibinfo {pages} {3988}
  (\bibinfo {year} {2011})}\BibitemShut {NoStop}%
\bibitem [{\citenamefont {Havriliak}\ and\ \citenamefont
  {Negami}(1966)}]{havriliak1966}%
  \BibitemOpen
  \bibfield  {author} {\bibinfo {author} {\bibfnamefont {S.}~\bibnamefont
  {Havriliak}}\ and\ \bibinfo {author} {\bibfnamefont {S.}~\bibnamefont
  {Negami}},\ }\href@noop {} {\bibfield  {journal} {\bibinfo  {journal}
  {{Journal of Polymer Science Part C-Polymer Symposium}}\ ,\ \bibinfo {pages}
  {{99+}}} (\bibinfo {year} {{1966}})}\BibitemShut {NoStop}%
\bibitem [{\citenamefont {Frohlich}(1958)}]{frohlich1958}%
  \BibitemOpen
  \bibfield  {author} {\bibinfo {author} {\bibnamefont {Frohlich}},\
  }\href@noop {} {\emph {\bibinfo {title} {Theory of Dielectics, Dielectric
  Donstant and Dielectric Loss}}}\ (\bibinfo  {publisher} {Oxford University
  Press, London},\ \bibinfo {year} {1958})\BibitemShut {NoStop}%
\bibitem [{\citenamefont {Huang}\ and\ \citenamefont
  {McKenna}(2001)}]{huang_new_2001}%
  \BibitemOpen
  \bibfield  {author} {\bibinfo {author} {\bibfnamefont {D.}~\bibnamefont
  {Huang}}\ and\ \bibinfo {author} {\bibfnamefont {G.~B.}\ \bibnamefont
  {McKenna}},\ }\href@noop {} {\bibfield  {journal} {\bibinfo  {journal} {The
  Journal of Chemical Physics}\ }\textbf {\bibinfo {volume} {114}},\ \bibinfo
  {pages} {5621} (\bibinfo {year} {2001})}\BibitemShut {NoStop}%
\bibitem [{\citenamefont {Struik}(1978)}]{Struik}%
  \BibitemOpen
  \bibfield  {author} {\bibinfo {author} {\bibfnamefont {L.~C.~E.}\
  \bibnamefont {Struik}},\ }\href@noop {} {\emph {\bibinfo {title} {Physical
  Aging in Amorphous Polymers and Other Materials}}}\ (\bibinfo  {publisher}
  {Elsevier, Amsterdam,},\ \bibinfo {year} {1978})\BibitemShut {NoStop}%
\bibitem [{\citenamefont {Yee}\ and\ \citenamefont {Smith}(1981)}]{yee1981}%
  \BibitemOpen
  \bibfield  {author} {\bibinfo {author} {\bibfnamefont {A.~F.}\ \bibnamefont
  {Yee}}\ and\ \bibinfo {author} {\bibfnamefont {S.~A.}\ \bibnamefont
  {Smith}},\ }\href@noop {} {\bibfield  {journal} {\bibinfo  {journal}
  {Macromolecules}\ }\textbf {\bibinfo {volume} {14}},\ \bibinfo {pages} {54}
  (\bibinfo {year} {1981})}\BibitemShut {NoStop}%
\bibitem [{\citenamefont {Hern\'{a}ndez}\ \emph {et~al.}(2013)\citenamefont
  {Hern\'{a}ndez}, \citenamefont {Sanz}, \citenamefont {Nogales}, \citenamefont
  {Ezquerra}, ,\ and\ \citenamefont {L\'{o}pez-Manchado}}]{rubber2013}%
  \BibitemOpen
  \bibfield  {author} {\bibinfo {author} {\bibfnamefont {M.}~\bibnamefont
  {Hern\'{a}ndez}}, \bibinfo {author} {\bibfnamefont {A.}~\bibnamefont {Sanz}},
  \bibinfo {author} {\bibfnamefont {A.}~\bibnamefont {Nogales}}, \bibinfo
  {author} {\bibfnamefont {T.~A.}\ \bibnamefont {Ezquerra}}, , \ and\ \bibinfo
  {author} {\bibfnamefont {M.}~\bibnamefont {L\'{o}pez-Manchado}},\ }\href@noop
  {} {\bibfield  {journal} {\bibinfo  {journal} {Macromolecules}\ ,\ \bibinfo
  {pages} {3176}} (\bibinfo {year} {2013})}\BibitemShut {NoStop}%
\bibitem [{\citenamefont {Lee}\ \emph {et~al.}(2009)\citenamefont {Lee},
  \citenamefont {Paeng}, \citenamefont {Swallen},\ and\ \citenamefont
  {Ediger}}]{lee2009}%
  \BibitemOpen
  \bibfield  {author} {\bibinfo {author} {\bibfnamefont {H.-N.}\ \bibnamefont
  {Lee}}, \bibinfo {author} {\bibfnamefont {K.}~\bibnamefont {Paeng}}, \bibinfo
  {author} {\bibfnamefont {S.~F.}\ \bibnamefont {Swallen}}, \ and\ \bibinfo
  {author} {\bibfnamefont {M.~D.}\ \bibnamefont {Ediger}},\ }\href@noop {}
  {\bibfield  {journal} {\bibinfo  {journal} {Science}\ }\textbf {\bibinfo
  {volume} {323}},\ \bibinfo {pages} {231 } (\bibinfo {year}
  {2009})}\BibitemShut {NoStop}%
\bibitem [{\citenamefont {Charvet}\ \emph {et~al.}(2019)\citenamefont
  {Charvet}, \citenamefont {Sotta}, \citenamefont {Vergelati},\ and\
  \citenamefont {Long}}]{charvet2019}%
  \BibitemOpen
  \bibfield  {author} {\bibinfo {author} {\bibfnamefont {A.}~\bibnamefont
  {Charvet}}, \bibinfo {author} {\bibfnamefont {P.}~\bibnamefont {Sotta}},
  \bibinfo {author} {\bibfnamefont {C.}~\bibnamefont {Vergelati}}, \ and\
  \bibinfo {author} {\bibfnamefont {D.~R.}\ \bibnamefont {Long}},\ }\href@noop
  {} {\bibfield  {journal} {\bibinfo  {journal} {Macromolecules}\ }\textbf
  {\bibinfo {volume} {52}},\ \bibinfo {pages} {6613} (\bibinfo {year}
  {2019})}\BibitemShut {NoStop}%
\bibitem [{\citenamefont {Djukic}(2020)}]{djukic2019}%
  \BibitemOpen
  \bibfield  {author} {\bibinfo {author} {\bibfnamefont {S.}~\bibnamefont
  {Djukic}},\ }\href@noop {} {\bibfield  {journal} {\bibinfo  {journal} {PhD
  thesis}\ } (\bibinfo {year} {to be defended in Lyon, 2020})}\BibitemShut
  {NoStop}%
\bibitem [{\citenamefont {Vogt}\ \emph {et~al.}(1990)\citenamefont {Vogt},
  \citenamefont {Dettenmaier}, \citenamefont {Spiess},\ and\ \citenamefont
  {Pietralla}}]{vogt1990}%
  \BibitemOpen
  \bibfield  {author} {\bibinfo {author} {\bibfnamefont {V.-D.}\ \bibnamefont
  {Vogt}}, \bibinfo {author} {\bibfnamefont {M.}~\bibnamefont {Dettenmaier}},
  \bibinfo {author} {\bibfnamefont {H.}~\bibnamefont {Spiess}}, \ and\ \bibinfo
  {author} {\bibfnamefont {M.}~\bibnamefont {Pietralla}},\ }\href@noop {}
  {\bibfield  {journal} {\bibinfo  {journal} {Colloid Polym. Sci.}\ }\textbf
  {\bibinfo {volume} {268}},\ \bibinfo {pages} {22} (\bibinfo {year}
  {1990})}\BibitemShut {NoStop}%
\end{thebibliography}%

\end{document}